\pdfoutput=1
\documentclass{llncs}
\usepackage{makeidx}  % allows for indexgeneration
\usepackage{graphicx}
\usepackage{commath}
\usepackage{amsmath}
\usepackage{amssymb}
\usepackage{caption}
\usepackage{multirow}
\usepackage{array}
\usepackage{blindtext}
\usepackage[english]{babel}
\usepackage{romannum}

\begin{document}

\frontmatter          % for the preliminaries
\pagestyle{headings}  % switches on printing of running heads
\pagenumbering{arabic}
\title{Hypothesis Generation\\
Using Link Prediction in a Bipartite Graph}
\titlerunning{Hypothesis Generation}  % abbreviated title (for running head)
\author{Jung-Hun Kim\inst{1} \and Aviv Segev\inst{2}}

% \author{}
%  \institute{}
  
%
\authorrunning{Jung-Hun Kim et al.} % abbreviated author list (for running head)

%%% list of authors for the TOC (use if author list has to be modified)
\tocauthor{Jung-Hun Kim, Aviv Segev}
\institute{
Graduate School of Knowledge Service Engineering\\
Department of Industrial and System Engineering\\ 
KAIST\\
Daejeon, South Korea\\
\email{junghunkim@kaist.ac.kr}\\ 
\and
CSAIL\\ MIT\\ Cambridge, MA USA\\
\email{aviv@csail.mit.edu}}

\maketitle              % typeset the title of the contribution

\begin{abstract}
% The process of generating novel ideas seems not to be comprehensible easily and naturally it is arduous for researchers to come up with new research ideas. For assisting researchers in this limitation,
The large volume of scientific publications is likely to have hidden knowledge that can be used for suggesting new research topics. We propose an automatic method that is helpful for generating research hypotheses in the field of physics using the massive number of physics journal publications. We convert the text data of titles and abstract sections in publications to a bipartite graph, extracting words of physical matter composed of chemical elements and extracting related keywords in the paper. The proposed method predicts the formation of new links between matter and keyword nodes based on collaborative filtering and matter popularity. The formation of links represents research hypotheses, as it suggests the new possible relationships between physical matter and keywords for physical properties or phenomena. The suggested method has better performance than existing methods for link prediction in the entire bipartite graph and the subgraph that contains only a specific keyword, such as `antiferromagnetism' or `superconductivity.'
% or NMR spectroscopy.
\end{abstract}
\section{Introduction}
The volume of scientific publications is growing at an exponential rate \cite{larsen2010rate}, which makes it impossible to keep up to date with all published papers. Automatic methods enabled by high-performance computing and big data mining algorithms can generate aggregate level insights that would not otherwise be uncovered by looking at data silos independently. We suggest a method for generating research hypotheses by extracting knowledge from massive amounts of published literature. Wallas \cite{wallas1926art} has suggested that generating new ideas is based on `Incubation,' which represents the subconscious without deliberate focus, and the `Illumination' phase, which represents a sudden flash of light. Because the process of generating ideas is vague, automated generation of hypotheses is a valuable tool that assists researchers in generating ideas. There is previous work that generated hypotheses automatically in biology using massive data from literature and experiments. We expand the field to physics, especially condensed matter physics, which deals with physical matter (e.g., Graphene, Silicon, FeSe). We adopt and improve the method for generating hypotheses based on the special characteristics of the field.

Condensed matter physics is one of the largest research fields in physics that deals with the physical properties of the phases of matter. In condensed matter physics, the researchers seek to understand the behaviors or properties of matter in various conditions, considering magnetization, electric field, mechanical stress, and temperature change. Also, they want to find the application of matter based on its properties. Some special behaviors or properties have a name like `superfluid,' `superconductivity,' `Bose-Einstein condensate (BCS)' or `antiferromagnetism,' which are considered as phenomena.
% There are some special techniques for analyzing matter like `NMR Spectrometer' or 'X-ray Spectroscopy.'
The relation between matter and phenomena is, for example, described as `The matter `YBa2Cu3O7' has High-temperature superconductivity phenomenon.' The study of such phenomena in matter is an interesting research topic and those phenomena are normally important keywords in the abstract section of papers. 

The proposed model suggests new research ideas in condensed matter physics based on relations between keywords and matter in the papers. Publications from 2004 to 2016 in the Physical Review B (PRB) journal 
%for condensed matter physics 
and the Physical Review Letter (PRL) journal, which are one of the representative journals for condensed matter physics,
%which are for the general field of physics
were used for the model. We extract matter only from the title and keywords from both the title and abstract of each article. We construct the bipartite graph using two types of nodes, matter and keywords, and edges or links which are formed when the matter and keywords appear in the title or abstract of the same article. A bipartite graph is a graph whose vertices can be divided into two disjoint and independent sets, where the sets refer to the matter and keywords sets in our graph, so that every edge connects a vertex in one set to the other set. 

Predicting the formation of new edges between nodes of matter and keywords represents that the two entities will co-occur in future literature in this research area. The new edges indicate the new relationships between matter and keywords and they contain new ideas which have not been considered previously. For predicting the formation of links in the bipartite graph, the proposed method uses collaborative filtering (CF) algorithms. Also, we found that the popularity of matter is an important factor for the formation of future links so we improve the CF algorithms considering the matter popularity from the appearance frequency in the publications.

Among the keyword nodes in the bipartite graph, we focus on `antiferromagnetism' and `superconductivity'. 
% and `NMR Spectroscopy' keyword nodes. 
Antiferromagnetism is one of the magnetic properties in matter. This magnetic property indicates that the magnetic moments of atoms or molecules align in the opposite direction of the spins of electrons and this property is applied to reading elements of hard-disk heads. 
Superconductivity is one of the hottest research topics in condensed matter physics because superconductivity is an interesting phenomenon of exactly zero electrical resistance. 
% and has many applications like MRI machines, magnetic fusion devices and maglev trains. Lastly, the NMR Spectrometer is the technique for analyzing physical and chemical properties of atoms or molecules. 
% It is widely used by chemists and biochemists to investigate the properties of organic molecules.
The keywords are reduced to `antiferromagnet' from `antiferromagnetic' or `antiferromagnetism' and `superconduct' from `superconductor(s),' `superconducting,' or `superconductivity' using stemming.
% and `NMR' from `NMR Spectroscopy' using stemming. 
The prediction of links between matter and those keyword nodes represents that we can predict matter that will be revealed to have a new relationship with those specific keywords. 

In this paper, we suggest a method for generating hypotheses in condensed matter physics and the method shows improved performance for predicting links in a bipartite graph in comparison with benchmark recommendation algorithms. It is the first time that automatic hypothesis generation is suggested in the field of physics using a massive amount of scientific literature. Our suggested concept for generating hypotheses can be easily extended to various other research topics.

% link prediction in a bipartite graph is used to generate research hypotheses automatically in the field of physics using a massive amount of scientific literature.

% The paper is organized as follows. Section \Romannum{2} describes the related work. Section \Romannum{3} describes the proposed method. Section \Romannum{4} includes the experiments, and Section \Romannum{5} shows the results. The last section, Section \Romannum{6}, summarizes the proposed method and suggests future works.
%
\section{Related Work}
%
%
% This section reviews the related works on link prediction and hypothesis generation.
% \subsection{Recommendation Algorithms}
% %
% Recommendation algorithms are widely used for recommending items to users in online services such as movies, songs, and products in markets like Netflix or Amazon. Collaborative filtering (CF) is a widely used technique in recommender systems, and its main concept is predicting the future preferences of a target user using the known preferences of other users. There are two types of collaborative filtering: memory-based and model-based. Memory-based algorithms use the patterns of users similar to the target user. Therefore, the similarity between users need to be calculated using cosine similarity or Pearson correlation \cite{resnick1994grouplens}. Considering the preferences of 10 users most similar to the target user, the target user's preference values for items are calculated and a system recommends items based on the values. The normally used model-based algorithm is the matrix factorization (MF) technique. 
% % There are several methods for MF: Singular Value Decomposition (SVD), Principal Component Analysis (PCA), Probabilistic Matrix Factorization (PMF), and Non-Negative Matrix Factorization (NMF) \cite{bokde2015matrix}. 
% Using MF, the approximated matrix to the original matrix consisting of the preference of users is calculated and the predicted values are obtained from the calculated matrix. The MF-based CF is known to work well for finding the hidden structure behind the data \cite{bokde2015matrix}. 

%
\subsection{Link Prediction in a Graph}
Link prediction in a graph is an active research area in computer science. Normally the type of graph is a unipartite graph such as a social network, web pages, and citation network. Liben-Nowelly \cite{liben2007link} suggested the idea for link prediction in the co-authorship network for predicting future interactions between researchers using measurements for network topologies.
% L$\ddot{u}$ \cite{lu2011link} suggested the methods for link prediction in a complex network using physical perspectives and approaches, such as random-walk-based methods and maximum likelihood methods. 
The recommendation problem can be seen as a link prediction in a bipartite graph.
In the case of link prediction for the bipartite graph, there is previous work using Collaborative  filtering (CF) algorithms, graph measures, and graph kernel-based machine learning \cite{li2013recommendation}. 

% However, We extend the applications of link predictions in a bipartite graph. Previous work focused on the domain of users selecting items like products or movies. Also, the work did not consider the characteristics of domains for improving the method of predicting links in a bipartite graph. 

%
\subsection{Hypothesis Generation}
There have been efforts in biology to make systems that generate research hypotheses by using text mining in the scientific literature of Medline abstracts or using algorithms for analyzing DNA data \cite{spangler2014automated,king2004functional}. Spangler \cite{spangler2014automated} constructed a system that can find the new protein kinases that phosphorylate the protein tumor suppressor p53 using graph-based diffusion of information. In genetics, 
% , a sequence motif is a nucleotide pattern and by using algorithms of motif-finding method, Hu \cite{hu2000combinatorial} analyzed the potential motif combination for hypotheses generation using DNA array data set.
King et al. \cite{king2004functional} applied a system to the determination of the gene function using deletion mutants of yeast which competes with human performance. However, previous works are normally limited to the field of biology and the methods are limited to a very specific purpose and hard to be generalized.\\\\
Our work extends the application of link prediction in a bipartite graph to generating hypotheses in physics and suggests an improved method for link prediction considering the characteristics of the domain. 
\section{Methodology}
\subsection{Construction of the Bipartite Graph}
%

% \begin{figure}[h!]
% \centering
% \includegraphics[scale=0.4]{bipatitegraph.png}
% \caption{Bipartite graph with matter and keyword nodes}
% \label{fig:bipartite graph}
% \end{figure}
%

For constructing matter nodes, we extract words of matter from the titles of publications but not from abstracts because we only consider the significant physical matter in each paper. First we remove the special characters (e.g.,`().-') in the title and then check whether each word is composed of the list of the chemical elements in the periodic table (e.g., Li, Ne, Ca) with some numbers, names of special materials, and some extra characters or notations. The following describes the text patterns used to extract words of matter in titles:
\begin{itemize}
  \item There is matter which is composed of the list of the chemical elements and numbers (e.g., TiSe2, Si(111), FeSe).
  \item There is matter which includes character `x' or `y' (e.g., BaFe2(As1-xPx)(2), FeTe1-xSex, InxGa1-xAs1-yNy).
  \item There is matter which includes some words `delta,' `beta,' `alpha,' `doped' and `based' (e.g., BiS2-based,  alpha-FeTe, beta-CaCr2O4).
  \item There is matter which includes notation `/' (e.g.,  Co/Cu, InAs/GaAs, Si/Ge).
  \item There are special materials which have a name themselves (e.g., graphene, silicone, diamond).
  \vspace{-20pt}
\end{itemize}
\begin{figure}[h!]
\scalebox{1.1}{
  \centering
  \begin{minipage}[b]{0.45\textwidth}
    \includegraphics[width=\textwidth]{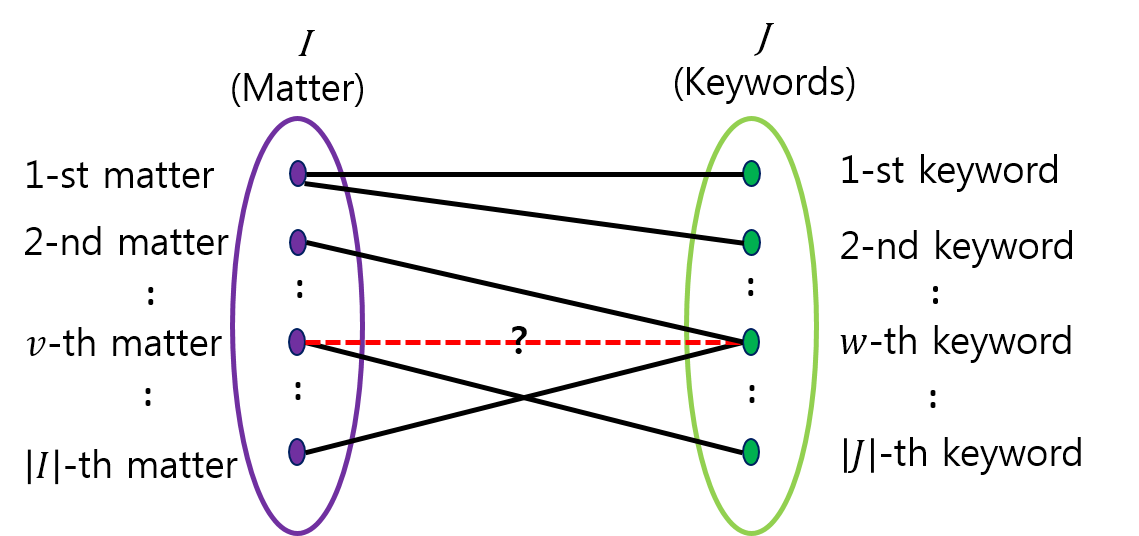}
\caption{Bipartite graph with matter and keyword nodes}
\label{fig:bipartite graph}
  \end{minipage}
  \hfill
  \begin{minipage}[b]{0.45\textwidth}
    \includegraphics[width=\textwidth]{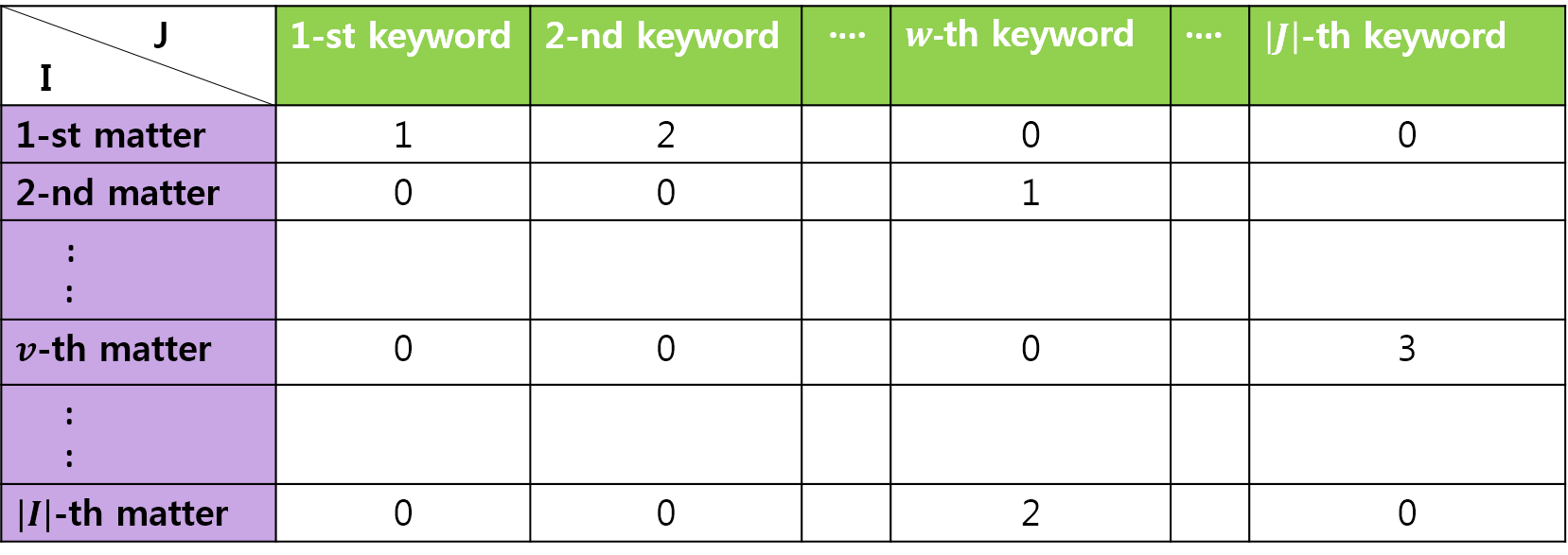}
   \caption{Example of an adjacency matrix $R$ using the graph in Fig. \ref{fig:bipartite graph}}
\label{fig:table}
  \end{minipage}}
  \vspace{-20pt}
\end{figure}
Lastly, we remove the words of matter whose length of the chemical element is one (e.g., O, N, S, and H) because normally the chemical elements with the length of one are very common elements and are less likely to have valuable meaning in the prediction. 

In the process of extracting keywords for constructing keyword nodes, first, we reduce each word to its root using stemming in each document composed of title and abstract. Then we use TF-IDF for each word and keep only the top 20 high TF-IDF valued keywords in each article, excluding the words of matter which cannot be in the keywords set of the bipartite graph. This allows us to select the important words in the paper as keywords which are likely to have a close and important relationship with the key matter in the title. As shown in Fig. \ref{fig:bipartite graph}, we construct a bipartite graph $G = (I + J, E)$ where $I$ is the set of matter nodes, $J$ is the set of keyword nodes, and $E$ is the set of edges between nodes in $I$ and $J$, which are formed when the two types of nodes appear in the same paper. 
%

% \begin{figure}[h!]
% \centering
% \includegraphics[scale=0.3]{table.png}
% \caption{Example of an adjacency matrix $R$ using the graph in Fig. \ref{fig:bipartite graph}}
% \label{fig:table}
% \end{figure}
% %

%

Its adjacency matrix $R\in \mathbb{R}^{|I|\times|J|}$ is defined as each element $r_{ij}=n$ where the matter of the $i$-th node and the keyword of the $j$-th node appear together in $n$ different publications for $i\in I$ and $j\in J$. Fig. \ref{fig:table} shows the example of an adjacency matrix $R$ using the graph in Fig. \ref{fig:bipartite graph}. To use the CF algorithms in our data set, we remove the matter which appears only once in the total publications to prevent the cold start problem \cite{su2009survey}, which occurs when a rare matter has less information in the graph. In the case of keyword nodes, we select the keywords which appear more than 100 times in the total publications and remove trivial keywords which only consist of numbers. 
\subsection{Link Prediction in the Bipartite Graph}
Collaborative filtering (CF) is used for movie recommendation in Netflix based on the user-item rating or the product recommendation to their customers in Amazon.com based on the purchase history. We consider the adjacency matrix $R$ in the bipartite graph as a user-item matrix for CF algorithms. By using CF algorithms in the matrix $R$ we can predict formations of new links that are not contained in the link set $E$ of the bipartite graph $G$ \cite{li2013recommendation}. We consider the matter nodes as users and the keyword nodes as items. 

For user-based CF which is one of the memory-based algorithms, we need to calculate the similarity between pairs of matter. We use cosine-based similarity (\ref{eq:1}) for all pairs of the matter in the set $I$ \cite{su2009survey}:
%1
\begin{equation} \label{eq:1}
sim(v_{1},v_{2})=cos(\vec{v_{1}},\vec{v_{2}})=\frac{\vec{v_{1}}\cdot\vec{v_{2}}}{\|\vec{v_{1}}\|_{2}\times\|\vec{v_{2}}\|_{2}}
\end{equation}
where $\vec{v_{1}}$, $\vec{v_{2}}$ are the $v_{1}$-th and $v_{2}$-th row vector in $R$ for $v_{1}$, $v_{2} \in I$, respectively. In the next step, let $v \in I$ and $w \in J$ for which value of element $r_{vw}$ in $R$ is zero. The zero value in the matrix $R$ represents that there is no link between the $v$-th matter and the $w$-th keyword. The following (\ref{eq:2}) is used when predicting the formation of new links with the user-based method \cite{su2009survey}:

%2
\begin{equation}  \label{eq:2}
\acute{r}_{vw}=\bar{r}_{v}+\frac{\sum_{\substack{u \in U_{m}}}(r_{uw}-\bar{r}_{u})\cdot sim(v,u)}{\sum_{\substack{u \in U_{m}}}|sim(v,u)|}
\end{equation}
where $\bar{r}_{v}$ is the average value of non-zero elements in the $v$-th row in $R$, the set $U_{m}$ is composed of the top-$m$ most similar matter to the target $v$-th matter among the entire matter using the similarity (\ref{eq:1}) and $\bar{r}_{u}$ is the average value of non-zero elements in the row of the matter $u \in U_{m}$ in $R$. The predicted value $\acute{r}_{vw}$ represents how likely the link is formed in the future so a higher value indicates a higher probability of the link formation. 

In the following Section 4, we show that the appearance frequency of matter words in publications, which represents the popularity of matter, is the critical factor for the appearance frequency of matter in the future research. Therefore, we suggest considering the popularity of matter by summation of the number of times it appears in the publication data, for both perspectives of negative and positive effects on the formation of links in the future. The modified predicted value $\dot{r}_{vw}$ considering user-based method and matter popularity (user-based MP) is (\ref{eq:3}, \ref{eq:4}):

%3 
\begin{equation}  \label{eq:3}
s_{vw}=\bar{r}_{v}+\frac{\sum_{\substack{u \in U_{m}^{*}}}(r_{uw}-\bar{r}_{u})\cdot sim(v,u)}{\sum_{\substack{u \in U_{m}^{*}}}|sim(v,u)|}
\end{equation}
%
%4
\begin{equation}  \label{eq:4}
\dot{r}_{vw}=\log({\sum_{\substack{j \in J}}r_{vj}})\times(s_{vw}+\alpha)
\end{equation}
where $U_{m}^*$ is the set composed of all elements in $U_{m}$, the top-$m$ most similar matter to the target $v$-th matter, and also the $v$-th matter itself. Instead of $U_{m}$, we use $U_{m}^*$ in (\ref{eq:3}) to consider the negative effect of matter popularity on the predicted value. For the negative effect, here is the explanation about the case when $u=v$ in the second term of (3). Note that the value of $r_{vw}$ is zero in the matrix $R$ and a larger $\bar{r}_{v}$ indicates that the $v$-th matter has more links, i.e. it is more popular. Therefore, if $\bar{r}_{v}$ is large, then the link formation between the $v$-th matter and the $w$-th keyword, which has not yet been formed, becomes a more rare event than the case when $\bar{r}_{v}$ is small. In other words, we can interpret the case when the $\bar{r}_{v}$ is large and $r_{vw}$ is zero as the formation of the specific link is a rare event, because there is no link between the $v$-th matter and the $w$-th keyword even though the $v$-th matter has been researched a lot. The value of $r_{vw}-\bar{r}_v$ which is negative in (\ref{eq:3}) represents how rarely the link will be formed between the $v$-th matter node and $w$-th keyword node and the value decreases the predicted value considering the rareness of the link formation.

% $r_{vw}$ is zero in the matrix $R$ and large $r_{vw}$ indicates that the $v$-th matter is popular and has many links. Therefore, if $r_{vw}$ is large, then the link between the $v$-th matter matter and the $w$-th keyword become a more rare event than the case when $r_{vw}$ is small. 

% where $U(m)^*$ is the set composed of $U(m)$, the top-$m$ most similar matter to the target $v$-th matter, and the $v$-th matter itself. Instead of $U(m)$, we use $U(m)^*$ to consider the negative effect of matter popularity on the predicted value. As $r_{vw}$ is zero in the matrix $R$, a larger $\bar{r}_{v}$ causes the link formation between the $v$-th matter and the $w$-th keyword to be more rare. The value of $r_{vw}-\bar{r}_v$ which is negative in (\ref{eq:3}) represents how rarely the link will be formed between the $v$-th matter node and $w$-th keyword node. Therefore, the rareness of link formation and the negative effect of matter popularity is considered by the predicted value $s_{vw}$ decreased by the negative value $r_{vw}-\bar{r}_v$.

% The value of $\sum_{\substack{j \in J}}r_{vj}$ is the summation of all values in the $v$-th row in matrix $R$ and represents the popularity of the $v$-th matter in the data.

On the other hand, $\log({\sum_{\substack{j \in J}}r_{vj}})$ in (\ref{eq:4}) is the weighting value for the positive effect of matter popularity. The value of $\sum_{\substack{j \in J}}r_{vj}$ is the summation of all values in the $v$-th row in matrix $R$ and represents the popularity of the $v$-th matter in the publications. The more popular the matter is the more likely it is to have new links. The role of constant $\alpha$ in (\ref{eq:4}) is to make all negative predicted values of $s_{vw}$ positive by positive parallel translation before they are weighted by the matter popularity. We sort the modified predicted values from user-based MP (\ref{eq:4}) in descending order. If the modified predicted value $\dot{r}_{vw}$ is high, the link has a higher probability to be formed in the future so the model recommends the links from the highest predicted valued link. 

Another memory-based algorithm is item-based CF, which is similar to the concept of user-based CF, except that it considers the similarity between items rather than users according to (\ref{eq:1}). The formula for item-based CF is as follows \cite{sarwar2001item}:

%5 
\begin{equation} \label{eq:5}
\acute{r}_{vw}=\bar{r}_{v}+\frac{\sum_{\substack{x \in X_{m}}}(r_{vx}-\bar{r}_{v})\cdot sim(v,x)}{\sum_{\substack{x \in X_{m}}}|sim(v,x)|}
\end{equation}
where the set $X_{m}$ is composed of the top $m$-most similar keywords to the $w$-th keyword.
% and $\bar{r}_{v}$ is the average value of $v$-th row in matrix $R$.
% instead of using $\bar{r}_{w}$ which is the average value of non-zero elements in the $w$-th column in $R$.
With the same perspective of the user-based algorithm, we suggest a new algorithm considering the negative and positive effect of the matter popularity for the item-based algorithm (item-based MP). The suggested formulas are as follows:

%6 
\begin{equation} \label{eq:6}
s_{vw}=\bar{r}_{v}+\frac{\sum_{\substack{x \in X_{m}^{*}}}(r_{vx}-\bar{r}_{v})\cdot sim(v,x)}{\sum_{\substack{x \in X_{m}^{*}}}|sim(v,x)|}
\end{equation}
\begin{equation} \label{eq:7}
\dot{r}_{vw}=\log({\sum_{\substack{j \in J}}r_{vj}})\times(s_{vw}+\sigma)
\end{equation}
where $X_{m}^*$ is the set composed of the $w$-th keyword and the elements in the set $X_{m}$. Equation (\ref{eq:6}) shows the predicted value considering the negative effect of matter popularity. In addition, the positive effect of matter popularity is considered in (\ref{eq:7}) with constant $\sigma$ that makes the negative values of $s_{vw}$ positive and this is the predicted value from item-based MP.

In the model-based algorithm for CF, we consider matrix factorization with matter popularity (MFMP). Let $P\in \mathbb{R}^{|I|\times K}$, $Q\in \mathbb{R}^{K\times|J|}$ be matrices with the parameter K of latent features number. The matrix factorization (MF) method is to find $ \widehat{R}=PQ$ which is the approximated matrix to the true adjacency matrix $R$ \cite{koren2009matrix}. Let the $i$-th row in $P$ be vector $\vec{p_{i}}$ and the $j$-th column in $Q$ be vector $\vec{q_{j}}$. In the MF method, the predicted value for the link between $v \in I$ and $w \in J$ is $\vec{q_{v}}^\mathsf{T}\vec{p_{w}}$. Considering the positive effect of matter popularity, the predicted value from MF is weighted by the matter popularity:

%8
\begin{equation} \label{eq:8}
\dot{r}_{vw}=\log({\sum_{\substack{j \in J}}r_{vj}})\times \vec{q_{v}}^\mathsf{T}\vec{p_{w}}
\end{equation}
Equation (\ref{eq:8}) gives the predicted value of the MFMP method. 
\\\\
In this section, we suggested three methods, user-based MP, item-based MP, and MFMP considering the matter popularity. In the next section, we compare the performance of the suggested methods with the existing methods. 
\section{Experiments}
\subsection{Datasets for the Recommendation System}

We use 45,603 publications in PRB and PRL from 2004 to 2012 as a training set and 15,624 publications from 2013 to 2016 as a test set for retrospective study. By setting the test set as the more recent data than the training set we can evaluate the performance of the concept of predicting the future links formation. After preprocessing the data as mentioned in Section 3, we get a $2807\times 1782$ matrix of $R$; the size of the matter set $I$ is 2,807 and the size of the keyword set $J$ is 1,782. 
\subsection{Distribution of Appearance Counts for Matter}
\begin{figure}[h!]
 \vspace{-10pt}
  \centering
  \begin{minipage}[b]{0.45\textwidth}
    \includegraphics[width=\textwidth]{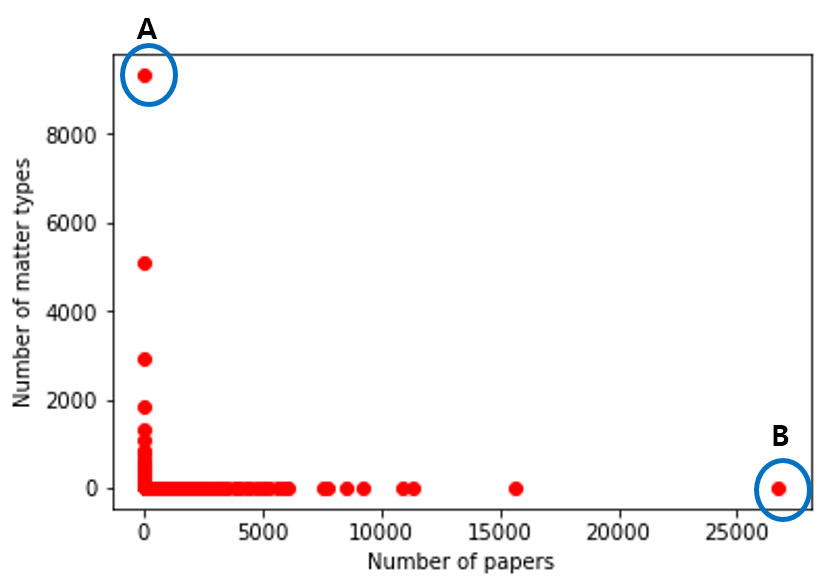}
    \vspace{-1em} 
    \caption{Distribution of appearance counts of matter in the total publications}
    \label{fig:distribution1}
  \end{minipage}
  \hfill
  \begin{minipage}[b]{0.45\textwidth}
    \includegraphics[width=\textwidth]{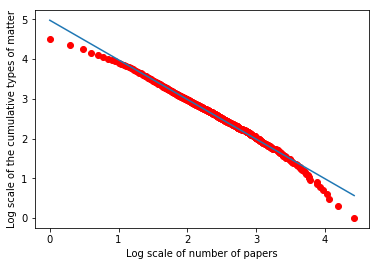}
    \vspace{-1em} 
    \caption{Log-log scale of cumulative distribution of appearance counts of matter in the total publications}
    
    \label{fig:distribution2}
  \end{minipage}
\end{figure}

We investigate the distribution of matter appearance counts in the titles and abstracts of the publications from 2000 to 2016.
The $y$-axis of Fig. \ref{fig:distribution1} indicates the number of different types of matter and the $x$-axis indicates the number of papers in which each matter appears. For better understanding the axes, we explain the point A and B in the plot. The point A represents that more than 8,000 different kinds of matter appear only once in the total papers and the point B represents that one kind of matter appears more than 25,000 times in the total papers. The plot in this figure follows the power law distribution.
For the more detailed investigation, the plot in the Fig. \ref{fig:distribution2} shows the log-log scale of cumulative distribution of the appearance counts of matter in the total publications following a straight line.
% Equations (\ref{eq:9}, \ref{eq:10}) are the cumulative distribution of power law distribution and the log-log scale of it.
%
%3
% \begin{figure}[h!]
% \centering
% \includegraphics[scale=0.4]{distribution1.png}
% \caption{Distribution of appearance counts of matter in the total publications}
% \label{fig:distribution1}
% \end{figure}
% 
%
% %4
% \begin{figure}[h!]
% \centering
% \includegraphics[scale=0.4]{distribution2.png}
% \caption{Log scale of cumulative distribution of appearance counts of matter in the total publications}
% \label{fig:distribution2}
% \end{figure}
% %
%
% %9
% \begin{equation} \label{eq:9}
% 1-F(x)=P(X\geq x)=\frac{c}{1-\gamma}x^{1-\gamma}
% \end{equation}
% %
% %10
% \begin{equation} \label{eq:10}
% \log P=C+(1-\gamma)\log(x)
% \end{equation}
% %
% and the fitted straight line has the slope of -0.9975. 
% It means the distribution is similar with the cumulative distribution of the power law distribution in log-log scale (\ref{eq:10}) with $\gamma=1.9975$ \cite{adamic2000zipf}.
From Fig. \ref{fig:distribution1} and Fig. \ref{fig:distribution2} we can say that the distribution of appearance counts of matter follows the power law distribution \cite{adamic2000zipf} and it shows that most publications are concentrated on only a few most popular types of matter.

% , and we can expect that this trend has continued and will continue like the well-known `the rich get richer' trend in wealth. The most frequently appearing matter is normally the single atom (e.g.,  Si, Fe, Co) and right after that the popular types of matter like `GaAs,' `diamond' and `graphite' follow.

%
\subsection{Benchmark Algorithms for Comparison}
We use the following methods for comparing the suggested algorithms: User-based MP, Item-based MP, and MFMP with the parameters $m$=10, $\alpha$=2.4, $\sigma$=0.01, and K=97.
% $\gamma$=0.0002, and $\lambda$=0.01.
\begin{enumerate}
  \item User-based: Simply use the predicted value from (\ref{eq:2}) \cite{su2009survey}.
  \item Item-based: Simply use the predicted value from (\ref{eq:5}) \cite{sarwar2001item}.
  \item Preferential Attachment: For a node $x$, we define $\Gamma(x)$ as the set of neighbors of $x$. A preferential attachment $|\Gamma(x)|\times|\Gamma(y)|$ recommends links according to the product of matter popularity and keyword popularity \cite{li2013recommendation}.
  %   \cite{chen2005link}.
  \item Matrix Factorization (MF): It is closely related to the singular value decomposition (SVD) and the predicted value is the element of the approximate matrix $\widehat{R}$ in the Section 3 \cite{li2013recommendation,koren2009matrix}.
  \item Random: Randomly choose the links for recommendation.
\end{enumerate}

\subsection{Investigation and Evaluation}
We investigate two different aspects of hypotheses generation using link prediction.
\begin{enumerate}
  \item We try to predict links in the range of the entire bipartite graph G. We compare the performance of user-based MP, item-based MP, and MFMP with five benchmark methods that we mentioned above. We evaluate each algorithm using the revised global receiver operating characteristic (GROC) curve which is slightly different from ROC or revised ROC curve \cite{li2013recommendation,schein2002methods}. In a GROC curve \cite{schein2002methods}, rather than evaluating performance by recommending the top-$k$ links for each matter, we evaluate the performance by recommending links from the entire graph between matter and keywords without limiting the number of recommendations in each matter. Therefore, the number of recommended links in each matter does not need to be the same. In the revised GROC \cite{li2013recommendation}, the $x$-axis is the number of recommendations rather than the false positive rate. The two variables are highly correlated so there is no great change in the shape of the curve and the revised curve is more straightforward for understanding the performance of the methods. After that, we plot the precision rate (\ref{eq:11}) for each method by increasing the recommendation number.
  \begin{equation} \label{eq:11}
  Precision=\frac{\scriptstyle{Number\ of\ recommended\ links \ that\ match\ with\ future \ links}}{\scriptstyle{Total\ number\ of\ recommended\ links}}
  \end{equation}
  In the practical aspect of the system, the precision is more important than recall because the experiments in physics for proving the recommendations from the system make a false positive rate costly. Therefore, precision represents how much the recommendations from each method are truthful. 
  \item We try to predict links between matter and the specific stemmed keyword `antiferromagnet' or `superconduct'. The area under ROC curve (AUROC) is calculated for both suggested methods and benchmark algorithms. After those evaluations, we investigate a more detailed example of the predicted links between matter and the keyword `antiferromagnet' using real text sentences in the publications
%   and `NMR' using real text sentences in the publications.
\end{enumerate}

\section{Results and Discussions}
\subsection{Link Prediction for Matter and Keywords}

\begin{figure}[h!]

  \centering
  \begin{minipage}[c]{0.4\textwidth}
   \centering
   \scalebox{1.1}{
   \includegraphics[width=\textwidth]{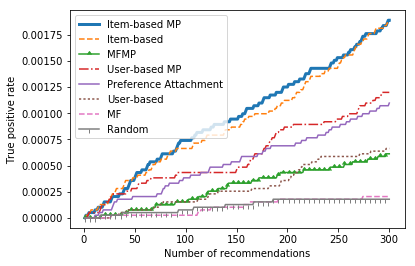}}
   
   \caption{Modified GROC results for the algorithms}
\label{fig:GROC for entire graph}
  \end{minipage}
  \hfill
  \begin{minipage}[c]{0.55\textwidth}
   \centering
   
    \includegraphics[width=\textwidth]{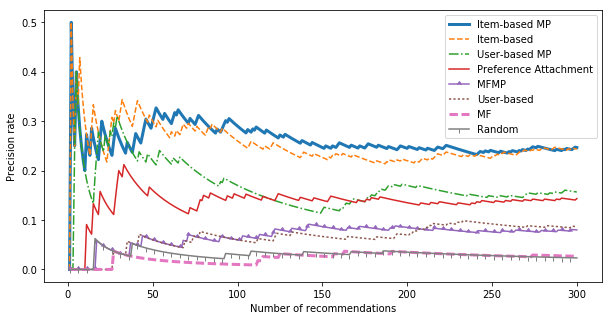}
    
   \caption{Precision rate for the algorithms}
\label{fig:Precision rate for entire graph}
  \end{minipage}
  \vspace{-10pt}
\end{figure}

For the first step, as mentioned in Section 4 we compare the performance of link predictions in the entire graph $G$ for suggested models and benchmark methods. Fig. \ref{fig:GROC for entire graph} shows the revised GROC curves of 300 recommendations.
%5
% \begin{figure}[h!]
% \centering
% \includegraphics[scale=0.5]{en_GROC.png}
% \caption{Modified GROC results for the algorithms}
% \label{fig:GROC for entire graph}
% \end{figure}
%
In Fig. \ref{fig:GROC for entire graph}, the item-based and item-based MP methods outperform the other methods, and the item-based MP is better than item-based in the range from about 100 to 250 recommendations.
% The suggested methods of item-based MP, user-based MP, and MFMP normally show better performance than the original methods of item-based, user-based and MF, respectively. Interestingly the memory-based algorithms, item-based and user-based, are better than the model-based algorithm, MF.
Fig. \ref{fig:Precision rate for entire graph} shows the precision rate for 300 recommendations. Normally the precision rate for all methods is the highest in the first 50 recommendations.
%6
% \begin{figure}[h!]
% \centering
% \includegraphics[scale=0.4]{en_precision.png}
% \caption{Precision rate for the algorithms}
% \label{fig:Precision rate for entire graph}
% \end{figure}
%
% After the point of the highest precision, as the number of recommendations increases, the precision rate decreases and converges because each method recommends links by the descending order of the predicted value, which represents the probability of the formation of the link. 
The item-based, item-based MP, and user-based MP methods show better precision rates than the other methods in the very first number of recommendations and item-based and item-based MP outperform the other methods in most of range. 

In the second step, as we mentioned in Section 4, instead of the entire graph $G$, we focus on the keywords `antiferromagnet' and `superconduct.' 
% and `NMR.'
The result of the experiments shows the performance of the link prediction between matter and the specific keywords. First, we did experiments for the `antiferromagnet' keyword. There are 2,360 zero elements among 2,807 elements in the column of the keyword `antiferromagnet' in the matrix $R$. This represents that there are 2,360 possible new links. In the test set, there are 44 new links for the keyword `antiferromagnet,' each representing a newly related matter with the keyword. The performance is the result of measuring how well each method recommends new links among 2,360 possible links for correctly predicting the 44 true links in the test set.
%%%
\begin{table}[h!]
% \vspace{-10pt}
\begin{minipage}[t]{0.45\textwidth}
\centering
\scalebox{0.8}{
    \begin{tabular}{ |c|c|c| } 
\hline
\textbf{Algorithms} & \textbf{AUROC} \\
 \hline
 MF & 0.5657 \\   
 MFMP & 0.6841 \\ 
 User-based & 0.6754 \\ 
 \textbf{User-based MP} & \textbf{0.7755} \\ 
 Item-based & 0.7418 \\ 
 \textbf{Item-based MP} & \textbf{0.7614} \\ 
 Preference Attachment & 0.6837 \\ 
 \hline
\end{tabular}}
\captionof{table}{AUROC for `antiferromagnet'} 
\label{table:AUROC_anti}
\end{minipage}%
\centering
\hspace{5ex}
\begin{minipage}[t]{0.45\textwidth}
\centering
\scalebox{0.8}{
\begin{tabular}{ |c|c|c| } 
\hline
\textbf{Algorithms} & \textbf{AUROC} \\
 \hline
 MF & 0.5821 \\   
 MFMP & 0.6327 \\ 
 User-based & 0.6962 \\ 
 \textbf{User-based MP} & \textbf{0.7350} \\ 
 Item-based & 0.5524 \\ 
 Item-based MP & 0.6199 \\ 
 Preference Attachment & 0.6303 \\ 
 \hline
\end{tabular}}
\captionof{table}{AUROC for `superconduct'}\label{table:AUROC_super}
\end{minipage}
% \begin{minipage}[t]{0.4\textwidth}
% \centering
% \scalebox{0.8}{
%     \begin{tabular}{ |c|c|c| } 
% \hline
% \textbf{Algorithms} & \textbf{AUROC} \\
%  \hline
%  MF & 0.6053 \\   
%  MFMP & 0.7181 \\ 
%  User-based & 0.5826 \\ 
%  \textbf{User-based MP} & \textbf{0.7678} \\ 
%  Item-based & 0.7099 \\ 
%  Item-based MP & 0.7254 \\ 
%  Preference Attachment & 0.7190 \\ 
%  \hline
% \end{tabular}}
% \captionof{table}{AUROC for `NMR'}\label{table:AUROC_NMR}
% \end{minipage}
\vspace{-30pt}
\end{table}
%
% \begin{table}[h!]
% \begin{center}
% \begin{tabular}{ |c|c|c| } 
% \hline
% \textbf{Algorithms} & \textbf{AUROC} \\
%  \hline
%  MF & 0.6053 \\   
%  MFMP & 0.7181 \\ 
%  User-based & 0.5826 \\ 
%  \textbf{User-based MP} & \textbf{0.7678} \\
%  Item-based & 0.7099 \\ 
%  Item-based MP & 0.7254 \\ 
%  Preference Attachment & 0.7190 \\ 
%  \hline
% \end{tabular}
% \captionof{table}{AUROC for `NMR'}\label{table:AUROC_NMR}
% \end{center}
% \end{table}
%
% %
% \begin{center}
% \begin{tabular}{ |c|c|c| } 
% \hline
% \textbf{Algorithms} & \textbf{AUROC} \\
%  \hline
%  MF & 0.5657 \\   
%  MFMP & 0.6841 \\ 
%  User-based & 0.6754 \\ 
%  \textbf{User-based MP} & \textbf{0.7755} \\ 
%  \textbf{Item-based} & \textbf{0.7418} \\ 
%  \textbf{Item-based MP} & \textbf{0.7614} \\ 
%  Preference Attachment & 0.6837 \\ 
%  \hline
% \end{tabular}
% \captionof{table}{AUROC for `antiferromagnet'} \label{table:AUROC_anti}
% \end{center}
% %
%7
\begin{figure}
% \vspace{-1em} 
\centering
\includegraphics[scale=0.4]{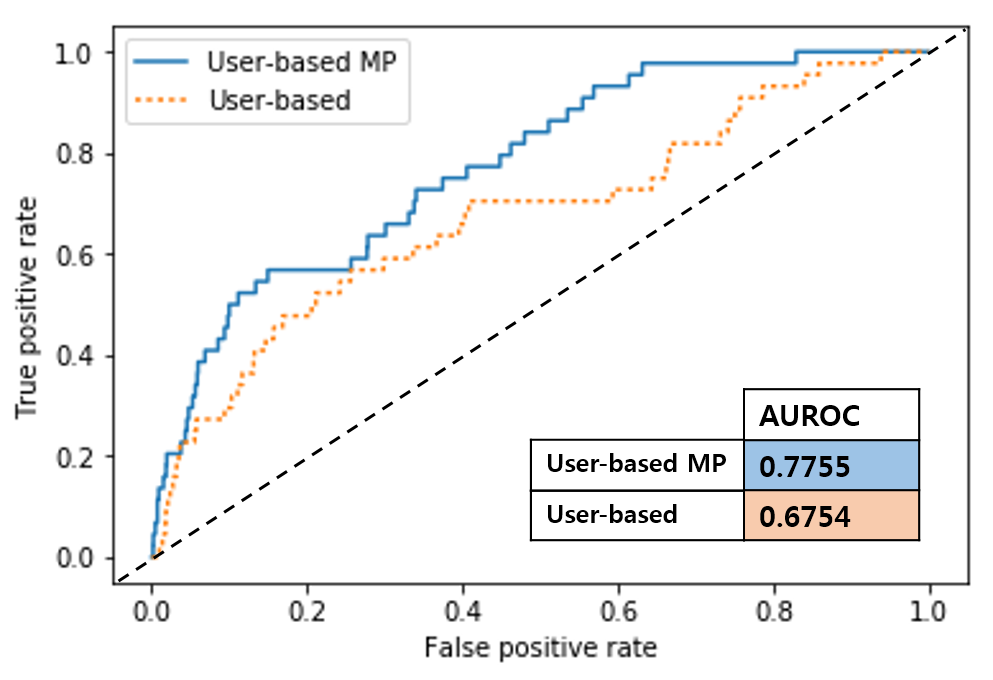}
 
\caption{ROC curves of user-based MP and user-based method for `antiferromagnet'}
\label{fig:ROC curve for anti}
% \vspace{-10pt}
\end{figure}

Table \ref{table:AUROC_anti} shows the AUROC value of each method for link prediction between matter and the `antiferromagnet' keyword. The bold values are the largest ones or are not significantly different from the largest one at 98\% confidence interval. The suggested methods, user-based MP and item-based MP, have better performance than the other methods.
In Fig. \ref{fig:ROC curve for anti}, we can see the improved performance of user-based MP compared to the original user-based method, with AUROC values of 0.7755 and 0.6754, respectively. The dashed line in the figure represents the performance of random method.
Table \ref{table:Real_anti} shows detailed examples of link prediction results for the keyword `antiferromagnet' within 100 recommendations. There is a total of 10 correct recommendations among 100. The predicted matter is always contained in the title but the keyword `antiferromagnet' is either in the title or abstract. The matter and keywords are likely to have a close relationship because the matter in the title is the key matter for the paper and the key matter is related to the keywords.
For the second keyword `superconduct,' there are 2,327 possible future links in the graph G and there are 33 true new links formed in the test set.
Table \ref{table:AUROC_super} shows the AUROC value of each method for link prediction between matter and the `superconduct' keyword. Based on the AUROC value, the suggested user-based MP algorithm outperforms other methods.
% Lastly, Table \ref{table:AUROC_NMR} shows the results of performance for the keyword `NMR.' We can see that user-based MP shows the best performance for `NMR.'
From the results of all the experiments for both specific keywords, user-based MP shows the best performance for all cases.
\begin{center}
 \vspace{4pt}
% \vspace{-1pt} 
\scalebox{0.7}{
\begin{tabular}{ | m{5em} | m{2cm}| m{8cm} | }
\hline
\textbf{Order in the list of recommendations}&\textbf{Matter} & \textbf{Title or abstract} \\ 
\hline
6 & Si & \textbf{Title}: \textbf{Antiferromagnetic} exchange interactions among dopant electrons in \textbf{Si} nanowires \\ 
\hline
8 & Eu & \textbf{Title}: Effect of \textbf{Eu} magnetism on the electronic properties of the candidate Dirac material EuMnBi2.
\textbf{Abstract}: Magnetic susceptibility measurements suggest \textbf{antiferromagnetic} (AFM) ordering of moments on divalent Eu ions near T-N = 22 K \\
\hline
15 & FeSe & \textbf{Title}: Spin Ferroquadrupolar Order in the Nematic Phase of \textbf{FeSe}.

\textbf{Abstract}: we find the FQ phase in close proximity to the columnar \textbf{antiferromagnet} commonly realized in iron-based superconductors.\\
\hline
22 & Fe-doped & Omitted\\
\hline
23 & Gd-doped & Omitted\\
% 22 & Fe-doped & \textbf{Title}: Magnetic phase transition in \textbf{Fe-doped} topological insulator Bi2Se3
% \textbf{Abstract}: For higher Fe concentration, >1.7 at. \%, Bi2Se3 prefers the \textbf{antiferromagnetic} phase mediated by the superexchange interaction. \\
% \hline
% 23 & Gd-doped & \textbf{Title}: Systematic study of the exchange interactions in \textbf{Gd-doped} GaN containing N interstitials, O interstitials, or Ga vacancies.
% \textbf{Abstract}: The exchange interactions between N interstitials (Ni) and Ni with Gd are found to be short-ranged and mainly \textbf{antiferromagnetic}.\\
\hline
28 & SrTiO3 & Omitted\\
\hline
43 & Cu(001) & Omitted\\
\hline
52 & Au & Omitted\\
\hline
55 & Fe1-xTe & Omitted\\
\hline
98 & Bi & Omitted\\
\hline
\end{tabular}}
% \vspace{-1em} 
\captionof{table}{Detailed investigation of correctly prediction results for `antiferromagnet' keyword within 100 recommendations} \label{table:Real_anti}
%\vspace{-1em} 
\end{center}

There are some limitations in this model. First, we consider only two journals, PRL and PRB, so even though there is no link in the training set or test set, the link can exist in other publications during the same period of our training set and test set, respectively. In addition, we only consider the period from 2004 to 2012 for the training set so the system does not have information of publications before 2004.
%
%
%
%
% \begin{center}
% \begin{tabular}{ |c|c|c| } 
% \hline
% \textbf{Algorithms} & \textbf{AUROC} \\
%  \hline
%  MF & 0.5821 \\   
%  MFMP & 0.6327 \\ 
%  User-based & 0.6962 \\ 
%  \textbf{User-based MP} & \textbf{0.7350} \\
%  Item-based & 0.5524 \\ 
%  Item-based MP & 0.6199 \\ 
%  Preference Attachment & 0.6303 \\ 
%  \hline
% \end{tabular}
% \captionof{table}{AUROC for `superconduct'}\label{table:AUROC_super}
% \end{center}
% %
%
% \begin{center}
% \begin{tabular}{ |c|c|c| } 
% \hline
% \textbf{Algorithms} & \textbf{AUROC} \\
%  \hline
%  MF & 0.6053 \\   
%  MFMP & 0.7181 \\ 
%  User-based & 0.5826 \\ 
%  \textbf{User-based MP} & \textbf{0.7678} \\
%  Item-based & 0.7099 \\ 
%  Item-based MP & 0.7254 \\ 
%  Preference Attachment & 0.7190 \\ 
%  \hline
% \end{tabular}
% \captionof{table}{AUROC for `NMR'}\label{table:AUROC_NMR}
% \end{center}
% %
% This causes a problem where the new links that have been recommended were in fact already formed before 2004 which is not included in our training set. Second, the method of the best performance is different depending on the purpose, whether it is for the entire graph or for the specific keyword.
% \vspace{-10pt}
%
\section{Conclusions and Future Works}
 \vspace{-2pt}
In this paper, we suggest a recommendation model for hypotheses generation in condensed matter physics. We convert the text data of publication into the bipartite graph using matter words and keywords. We propose methods for predicting links in the graph using CF algorithms with matter popularity: user-based MP, item-based MP, and MFMP. We can gain better performance in our suggested methods for both cases, the entire graph and the subgraphs. From the results of the subgraphs, we confirm that our model can be applied to other various keywords depending on the purpose of the research. 

Future works include expanding the data set. Also, we can try to apply our method to other subgraphs for specific purposes such as the keywords `superfluid,' `BCS' and `ferromagnetism,' which are important concepts in condensed matter physics. Lastly, we can try to modify the prediction model using the graphical model approach.
\vspace{-10pt}
\medskip

\bibliographystyle{splncs03}
\bibliography{mybib}

\begin{thebibliography}{10}
\providecommand{\url}[1]{\texttt{#1}}
\providecommand{\urlprefix}{URL }

\bibitem{adamic2000zipf}
Adamic, L.A.: Zipf, power-laws, and pareto-a ranking tutorial. Xerox Palo Alto
  Research Center, Palo Alto, CA, http://ginger. hpl. hp.
  com/shl/papers/ranking/ranking. html  (2000)

\bibitem{king2004functional}
King, R.D., Whelan, K.E., Jones, F.M., Reiser, P.G., Bryant, C.H., Muggleton,
  S.H., Kell, D.B., Oliver, S.G.: Functional genomic hypothesis generation and
  experimentation by a robot scientist. Nature  427(6971),  247--252 (2004)

\bibitem{koren2009matrix}
Koren, Y., Bell, R., Volinsky, C.: Matrix factorization techniques for
  recommender systems. Computer  42(8) (2009)

\bibitem{larsen2010rate}
Larsen, P.O., Von~Ins, M.: The rate of growth in scientific publication and the
  decline in coverage provided by science citation index. Scientometrics
  84(3),  575--603 (2010)

\bibitem{li2013recommendation}
Li, X., Chen, H.: Recommendation as link prediction in bipartite graphs: A
  graph kernel-based machine learning approach. Decision Support Systems
  54(2),  880--890 (2013)

\bibitem{liben2007link}
Liben-Nowell, D., Kleinberg, J.: The link-prediction problem for social
  networks. journal of the Association for Information Science and Technology
  58(7),  1019--1031 (2007)

\bibitem{sarwar2001item}
Sarwar, B., Karypis, G., Konstan, J., Riedl, J.: Item-based collaborative
  filtering recommendation algorithms. In: Proceedings of the 10th
  international conference on World Wide Web. pp. 285--295. ACM (2001)

\bibitem{schein2002methods}
Schein, A.I., Popescul, A., Ungar, L.H., Pennock, D.M.: Methods and metrics for
  cold-start recommendations. In: Proceedings of the 25th annual international
  ACM SIGIR conference on Research and development in information retrieval.
  pp. 253--260. ACM (2002)

\bibitem{spangler2014automated}
Spangler, S., Wilkins, A.D., Bachman, B.J., Nagarajan, M., Dayaram, T., Haas,
  P., Regenbogen, S., Pickering, C.R., Comer, A., Myers, J.N., et~al.:
  Automated hypothesis generation based on mining scientific literature. In:
  Proceedings of the 20th ACM SIGKDD international conference on Knowledge
  discovery and data mining. pp. 1877--1886. ACM (2014)

\bibitem{su2009survey}
Su, X., Khoshgoftaar, T.M.: A survey of collaborative filtering techniques.
  Advances in artificial intelligence  2009, ~4 (2009)

\bibitem{wallas1926art}
Wallas, G.: The art of thought harcourt. Bruce and Company, New York  (1926)

\end{thebibliography}

\end{document}